\documentclass[aps,twocolumn,a4paper,showpacs]{revtex4}
\usepackage{graphicx}
\usepackage{amsmath}
\usepackage{amssymb}
\usepackage{enumerate}
\usepackage{subfigure}
\usepackage{tabularx}
\newcommand{\be}{\begin{equation}}
\newcommand{\ee}{\end{equation}}
\newcommand{\ben}{\begin{eqnarray}}
\newcommand{\een}{\end{eqnarray}}
\newcommand{\bes}{\begin{subequations}}
\newcommand{\ees}{\end{subequations}}

\newcommand{\bb}{\bibitem}
\newcommand{\sech}{{\rm sech}}

\begin{document}
\title{Models for asymmetric hybrid brane}
\author{D. Bazeia$^{1}$, M.A. Marques$^1$, and R. Menezes$^{2,3}$}
\affiliation{$^1$Departamento de F\'\i sica, Universidade Federal da Para\'\i ba, 58051-970 Jo\~ao Pessoa, PB, Brazil}
\affiliation{$^2$Departamento de Ci\^encias Exatas, Universidade Federal da Para\'{\i}ba, 58297-000 Rio Tinto, PB, Brazil}
\affiliation{$^3$Departamento de F\'\i sica, Universidade Federal de Campina Grande, 58109-970, Campina Grande, PB, Brazil}
\begin{abstract}
We deal with relativistic models described by a single real scalar field, searching for topological structures that behave asymmetrically, connecting minima with distinct profile. We use such features to build a new braneworld scenario, in which the source scalar field contributes to generate asymmetric hybrid brane.
\end{abstract}
\date{\today}
\maketitle

\section{Introduction}

The braneworld concept which appeared in \cite{rs} deals with a warped geometry and engenders a single extra spatial dimension of infinite extent. It gives rise to the thin brane profile, but the scenario was soon modified to support the thick brane profile  suggested in \cite{gw} and further investigated in \cite{tb1,tb2,tb3,tb4,tb5,tb6,tb7,tb8,tb9,tb10}, as well as in many other more recent works. In the thick brane scenario, the brane appears through the inclusion of a source scalar field, which, in the absence of gravity, is capable of supporting topological structure known as kink. The scenario is such that when one embeds the source scalar field into the Einstein-Hilbert action with the warped geometry with a single extra dimension of infinite extent, the topological structure that appears from the scalar field generates the thick brane configuration.

The thick brane scenario in general engenders a symmetric brane, since the source scalar field model presents parity or $Z_2$ symmetry. This means that the profile of the brane along the extra dimension is the same, at both the left and right side. However, one can also consider an asymmetric brane, if the scalar field model can support an asymmetric structure. This means that the profile in the left side of the brane is different from that of the right side. The asymmetric feature of a brane is of current interest, and has been studied by several authors with distinct motivation in \cite{a0,a1,a2,a3,a4,a5,a6,a7,a8,a10,a10a,a11,a12}. 

An important effect of the asymmetric structure of the brane is to contribute to accelerate the Universe, if one investigates cosmic evolution in a braneworld scenario constructed on top of an asymmetric configuration. As we shall see in the next section, the asymmetry of the potential makes the solution and the energy density asymmetric, and this modifies its behavior. In particular, the force \cite{M,T} the asymmetric configuration engenders is also asymmetric, and may induces acceleration in the braneworld cosmology constructed from an asymmetric structure, or in a braneworld scenario similar to the one investigated in \cite{brito}.

In the recent work \cite{hb} one has offered a new type of braneworld scenario, with the brane engendering a hybrid behavior. This is possible when the source scalar field supports localized structure with compactlike profile \cite{hb}. In this case, however, the hybrid brane configuration is symmetric, so we refer to it as the symmetric hybrid brane profile. An interesting issue which is the purpose of the current work is then to extend the hybrid brane profile in order to include the case of an asymmetric hybrid brane, which allows to explore possibilities that are not present in the symmetric case. 

The issue is nontrivial, because the presence of compactlike structures requires differential equations that are more complicated, when compared to the case of the standard kinks. For this reason, we first focus on the construction of an asymmetric compactlike structure. We follow the route introduced in \cite{hb}, and below we investigate the presence of  compactlike structures that are asymmetric. In Sec.~\ref{sec:brane} we embed a scalar field source model in a warped geometry with a single extra spatial dimension of infinite extent, and we build the asymmetric braneworld scenario. We end the work in Sec.~\ref{sec:end}, where we summarize the results and add some directions for further study. 

\section{Asymmetric structures}

In order to investigate the problem, we start from a Lagrange density with standard kinematics, describing a real scalar field $\phi$. We take the metric $\text{diag}(+,-)$, and use $V(\phi)$ as the potential that identifies the way the scalar field self-interacts. 
The subject was already investigated in \cite{hb}, so we briefly review the main results as follows.
We work with dimensionless field and coordinates, and the equation of motion for static solution is
\be\label{eqMotion1}
\frac{d^2 \phi}{dx^2}  = \frac{dV}{d\phi}.
\ee
We suppose that $V(\phi)$ supports at least the two minima $\{v_i, i=1,2\},$ with $dV/d\phi|_{v_i}=0$ and $V(v_i)=0$.  A topological sector is characterized by two neighbor minima, and the topological solution is a kinklike configuration, $\phi(x)$, that connects the minima $v_1$ and $v_2$, such that $\phi(x\to -\infty)\to v_1$ and $\phi(x\to\infty)\to v_2$, with vanishing derivatives. 

The kinklike solution goes asymptotically to the minima, depending exponentially on the mass associated to the corresponding minima. Thus, for larger and larger values of the mass, the exponential reaches the minimum faster and faster, and in the limit $m\to \infty$, the solution tends to become a compact configuration. The solution has energy density $\rho(x)$, given by 
\be
\rho(x)\!=\!\frac12 \left(\!\frac{d\phi}{dx}\!\right)^2\!\!\!+V(\phi(x))\!=\!\left(\!\frac{d\phi}{dx}\!\right)^2\!\!\!=\!2V(\phi(x)).
\ee

We study linear stability, adding small fluctuations around the static solution $\phi(x)$, writing $\phi(x,t)=\phi(x)+\eta(x)\cos(\omega t)$. We use this into the equation of motion and expand it up to first-order in $\eta$ to get the Schroedinger-like equation 
\be \label{stab}
\left(-\frac{d}{dx^2}+U(x) \right)\eta=\omega^2 \eta, \,\,\,\,  U(x)=\left.\frac{d^2V}{d\phi^2}\right|_{\phi=\phi(x)}.
\ee 
We see from the above Eq.~(\ref{stab}) that the stability potential goes asymptotically to $m^2_1$ and $m^2_2$, which are the (squared) masses of the elementary excitations at the minima $v_1$ and $v_2$, respectively.

A special case is the one in which the potential can be written as
\be\label{pot}
V=\frac12\left(\frac{dW}{d\phi}\right)^2=\frac12 W_\phi^2,
\ee
where $W=W(\phi)$. It provides a first order differential equation for the field
\be\label{fo}
\frac{d\phi}{dx} = \frac{dW}{d\phi},
\ee
and the energy density can be written as
\be
\rho(x)=\frac{dW}{dx}.
\ee
The total energy of the kinklike structure is then 
\be
E = \left|W(v_{2}) - W(v_1)\right|.
\ee
A model of importance is the well-known $\phi^4$ model, with spontaneous symmetry breaking. It is described by the function
\be
W(\phi) = \phi-\frac13{\phi^3}.
\ee
Thus, the potential is
\be\label{phi4}
V(\phi) = \frac12 (1-\phi^2)^2.
\ee
The minima are at $v_\pm = \pm1$ and the maximum is at the origin, with $V(0)=1/2$. The masses are $m_\pm^2 = 4$ and the equation of motion is
\be\label{eqmotionphi4}
\frac{d^2\phi}{dx^2}=-2\phi(1-\phi^2)
\ee
whose solution is
\be
\phi(x) = \tanh(x),
\ee
which we have centered at the origin. The energy density takes the form
\be
\rho(x)=\sech^4(x)
\ee
and the energy is $E=4/3$. The stability potential is
\be
U(x)=4-6\,\sech^4(x).
\ee
The topological kinklike structure is linearly stable. In fact, when the potential is written as in (\ref{pot}), the solution obeys the first-order Eq.~(\ref{fo}) and we can write the stability potential of (\ref{stab}) as
\be
U(x)=W^2_{\phi\phi}+W_\phi W_{\phi\phi\phi},
\ee
and so the Schr\"odingerlike operator in (\ref{stab}) can be factored in the form
\be
H=S^\dag S;\;\;\;S=\frac{d}{dx}-W_{\phi\phi}.
\ee
This operator is non negative, so there is no negative eigenvalue, thus ensuring linear stability.

This model engenders the $Z_2$ symmetry, and so it gives rise to a symmetric structure. 
However, we want to study asymmetric brane, so we need to build asymmetric kinklike structures. We get inspiration from the previous study \cite{hb} and below we investigate two distinct models, where the asymmetry is controlled by a single parameter, which modifies the mass of the field configuration, making the force \cite{M,T} the solution engenders asymmetric. This provides a new mechanism to induce acceleration in the appropriate braneworld scenario, as the one investigated before in Ref.~\cite{brito}.

\subsection{Model 1} 
We start with the model described by the following function
\be\label{superpot2}
W(\phi) = \phi -\frac{\phi^2}{2} + \frac{\phi^{1+p}}{1+p} - \frac{\phi^{2+p}}{2+p},
\ee
which leads to the potential
\be\label{pot2}
V(\phi)=\frac12 (1-\phi)^2\left(1+\phi^p\right)^2,
\ee
where $p$ is odd integer, $p=1,3,5\ldots$; we note that it reproduces the $\phi^4$ model for $p=1$. This potential has the minima $\phi_\pm=\pm1$ and the associated masses $m^2_{+}=4$ and $m^2_{-}=4p^2$. The parameter $p$ induces the required asymmetry.
The maximum of the potential in between the two minima can be found through the algebraic equation 
\be\label{maximumpointsmodel2}
1 -p{\phi}^{p-1} + p\left( p+1 \right) {\phi}^{p}=0.
\ee
We see that it is at $\phi=0$ for $p=1$.

\begin{figure}[t]
\includegraphics[width=4.2cm]{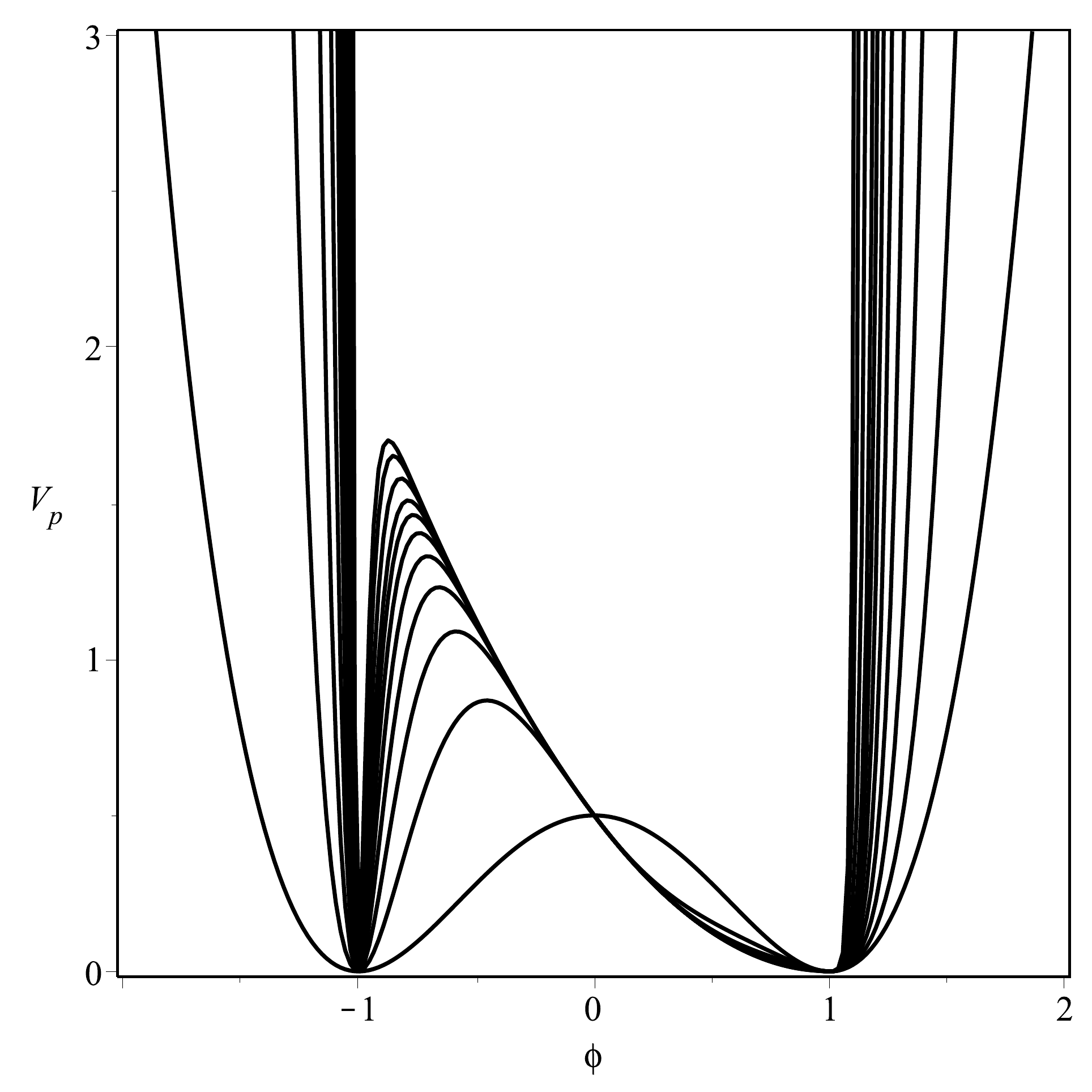}
\includegraphics[width=4.2cm]{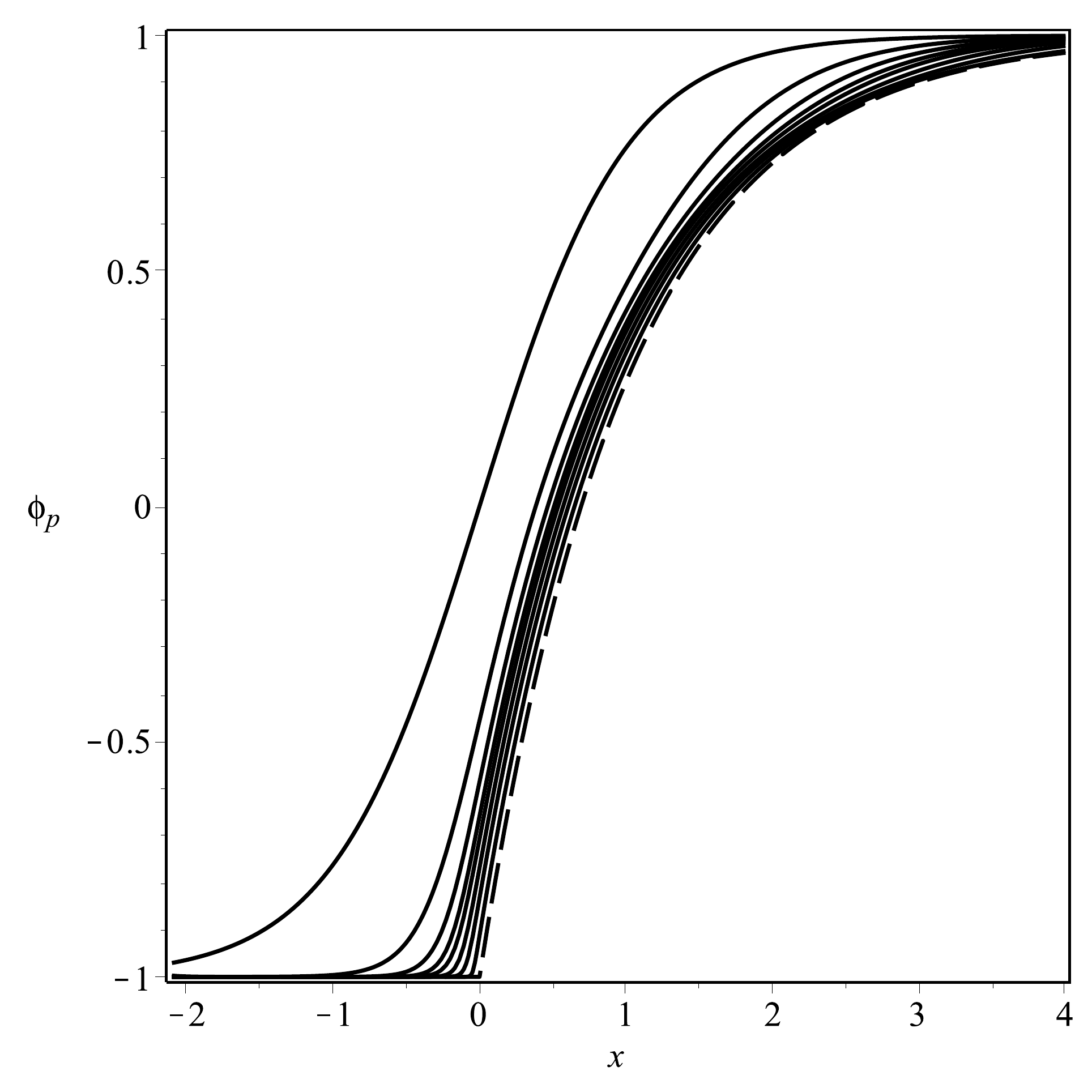}
\includegraphics[width=4.2cm]{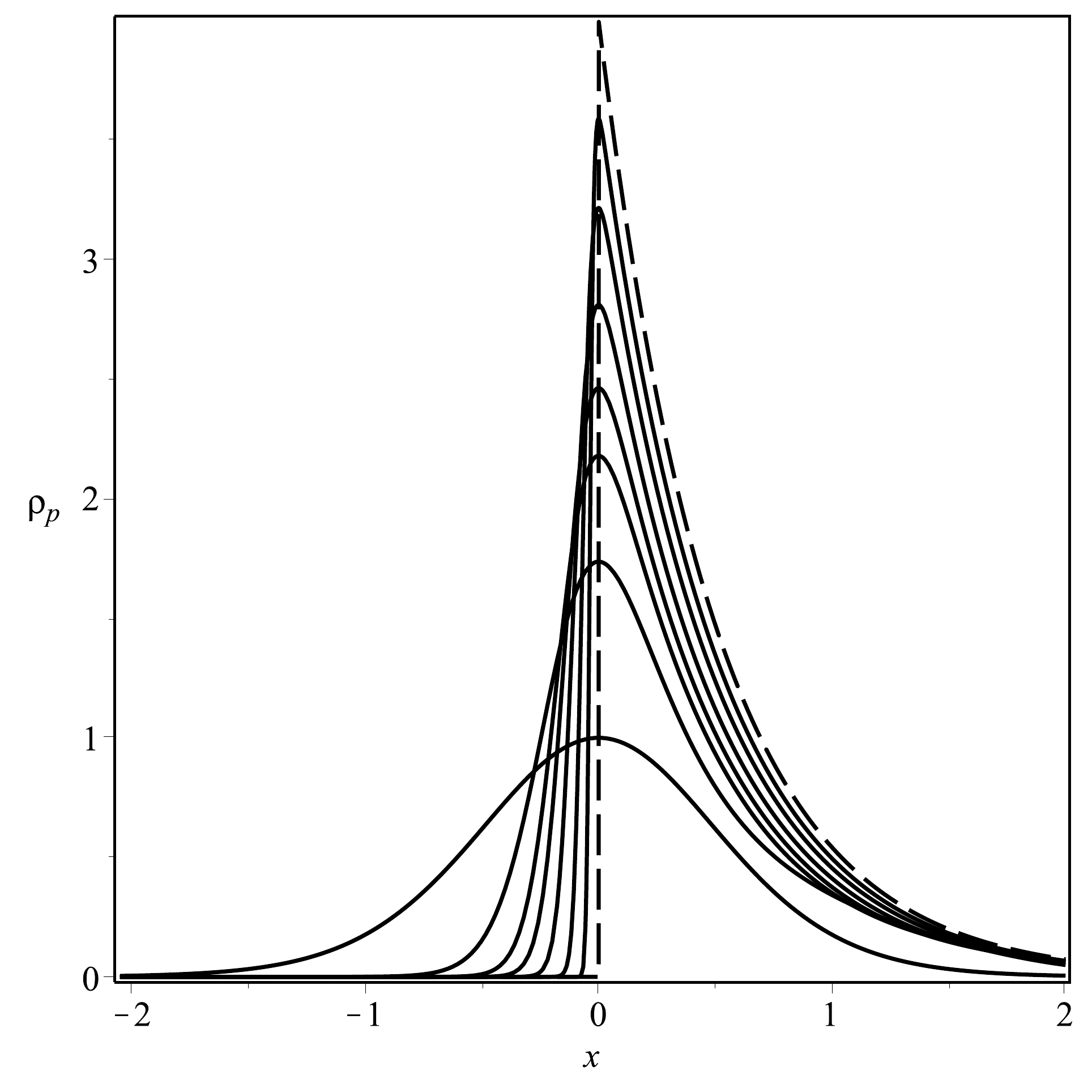}
\includegraphics[width=4.2cm]{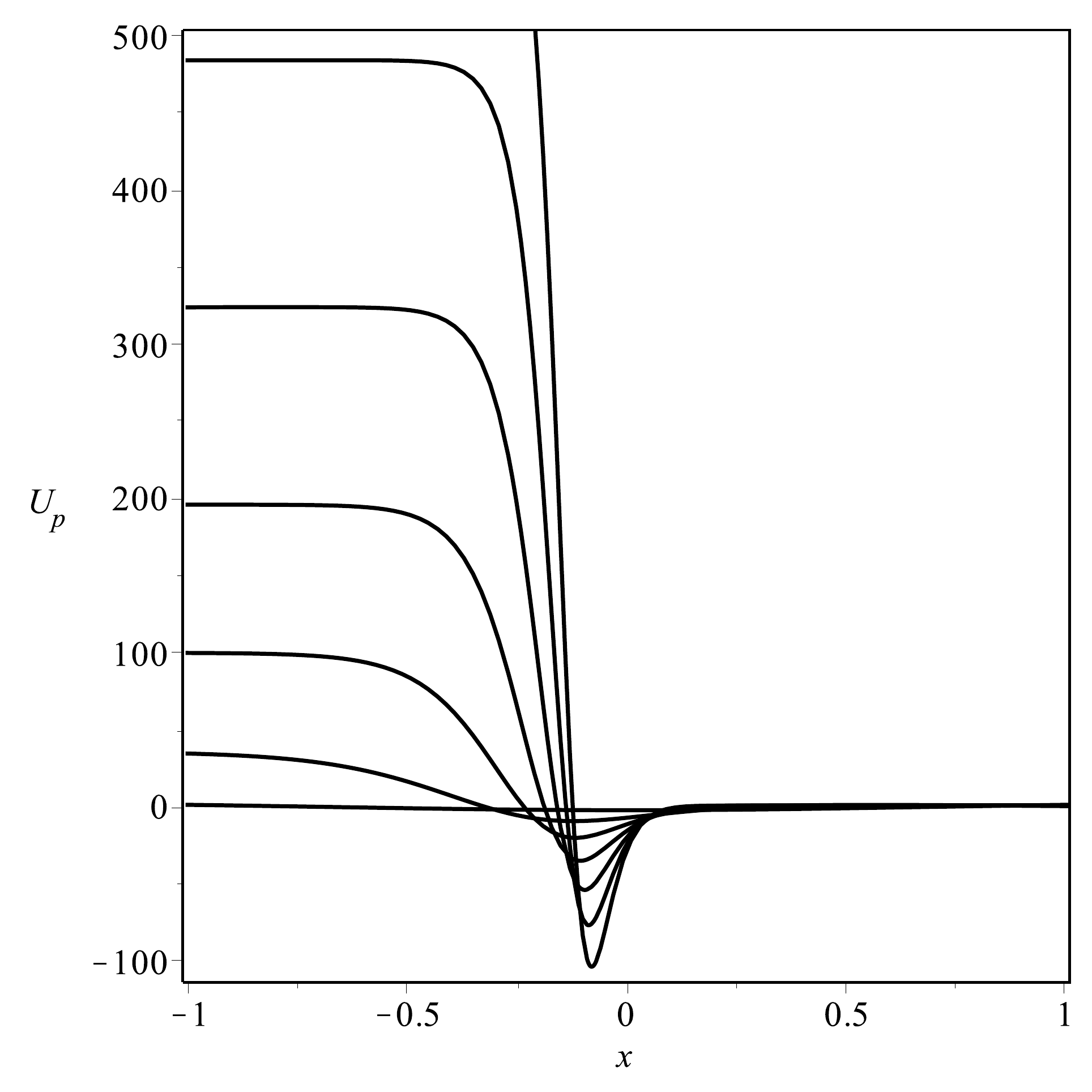}
\caption{The potential (\ref{pot2}) (top left), the kinklike solution (top right), the energy density (bottom left) and the stability potential (bottom right) depicted for several values of $p$. The dashed lines stand for the case $p\to\infty$.}
\label{fig3}
\end{figure}

This model engenders kinklike solution that obeys first-order differential equation, so it is linearly stable.
We have investigated the model numerically, and in Fig.~\ref{fig3} we depict 
the potential, kinklike solution, energy density and the stability potential for several values of $p$. The energy is given by 
\be\label{enep}
E=\frac{2+2p}{2+p}.
\ee
As expected, for $p=1$ we get the energy of the $\phi^4$ model, that is, $E=4/3$. As we take larger and larger values of $p$, the energy tends to $E\to2$.

In the limit $p\to\infty$, the solution becomes
\be
\phi(x)=
\begin{cases}
-1, \,\,\, & x<0,\\
1-2e^{-x},\,\,\,& x\geq0.
\end{cases}
\ee 
The energy density is
\be
\rho(x)=
\begin{cases}
0, \,\,\, & x< 0,\\
4e^{-2x},\,\,\,& x\geq 0.
\end{cases}
\ee
After integration, we note that the energy becomes $E=2$, which agrees with the previous result, below Eq.~(\ref{enep}). Also, the stability potential has the form
\be
U(x)=
\begin{cases}
\infty, \,\,\, & x<0,\\
-\infty, \,\,\, &x=0,\\
1,\,\,\,& x>0.
\end{cases}
\ee 

\begin{figure}[t]
\includegraphics[width=4.2cm]{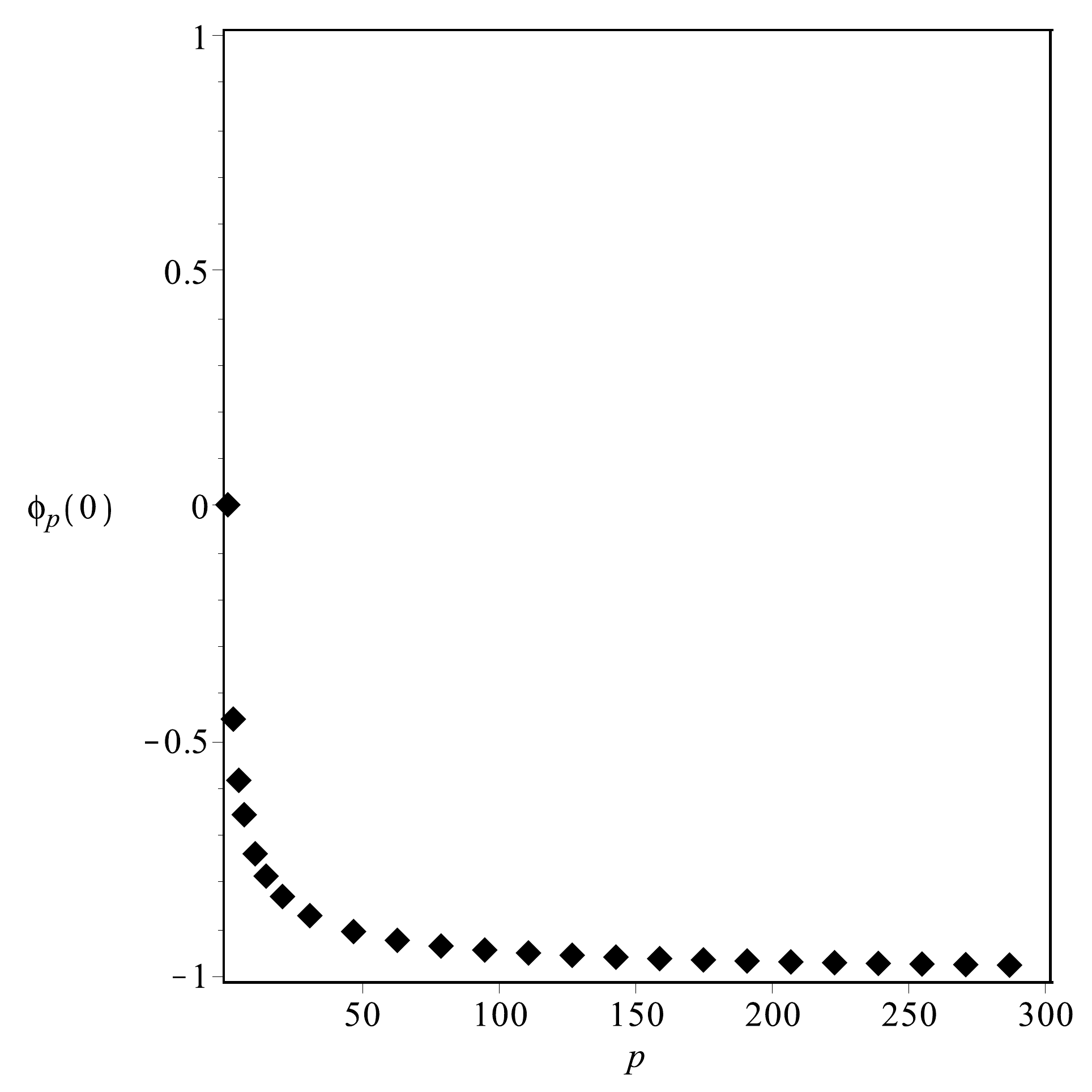}
\includegraphics[width=4.2cm]{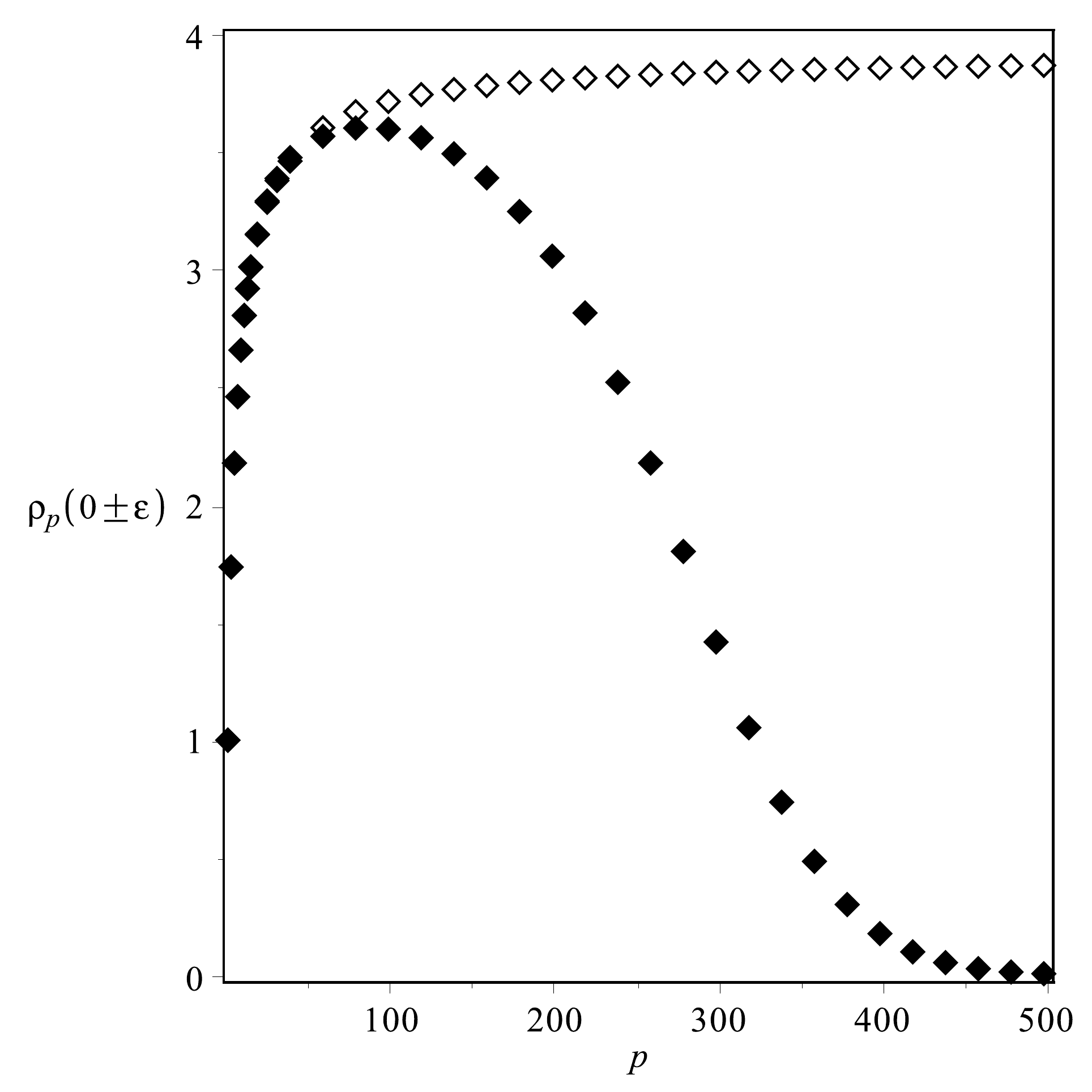}
\caption{The solution $\phi(0)$ (left) and the energy density $\rho(x=\pm0.01)$ (right), depicted for several values of $p$. In the right picture, the solid diamonds are for $x=-0.01$, and the hollow diamonds for $x=0.01$.}
\label{fig4}
\end{figure}

In Fig.~\ref{fig4} we depict the solution $\phi(0)$ and the energy density at $x=0\pm0.01$. We note that the plots for $\rho(0\pm0.01)$ agree with the general behavior, and shows a discontinuity as $p$ increases to larger and larger values.

\subsection{Model 2}

Now, we introduce a model that can be solved analytically. It reminds the $\phi^6$ model, and has the following superpotential
\be\label{superpotphi6like}
W_n(\phi)=\frac12 n\phi^2 - \frac12 \frac{n}{n+1}\phi^{\frac{2(n+1)}{n}},
\ee
and potential
\be\label{potphi6like}
V_n(\phi)=\frac12 n^2\phi^2\left(1-\phi^{\frac2n}\right)^2.
\ee
Here $n$ is a positive integer. In fact, we get the $\phi^6$ model for $n=1$, that is,
\be  
W_1 = \frac12 \phi^2-\frac14\phi^4, \;\;\;\;\;\; V_1 = \frac12 \phi^2(1-\phi^2)^2,
\ee
and in the limit $n\to\infty$,
\be
W_\infty = \frac12 \phi^2(1-\ln(\phi^2)), \;\;\;\;\;\; V_\infty = \frac12 \phi^2\ln^2(\phi^2).
\ee
The potential for $n$ arbitrary has the minima in $\phi_\pm=\pm1$ and $\phi_0=0$ for any value of $n$. The (classical, squared) mass of the scalar field in each minimum is $m_+^2=m_-^2=4$ and $m^2_0=n^2$. Thus, as we increase $n$, $m_0$ also increases and the solution tends to become a half-compacton. The maxima in between the three minima are
\be\label{phimaxanalytic}
\phi^\pm_{ m_n}=\pm\left(\frac{n}{n+2}\right)^{\frac n2}.
\ee
Therefore, the maximum value of the potential in between the minima is $V(\phi^\pm_{m_n})=2\left(n/(n+2)\right)^{n+2}$, and we see that it is finite in the limit $n\to\infty$: $\phi^\pm_{m_\infty}=\pm e^{-1}$, which gives $V(\phi^\pm_{m_\infty})=2e^{-2}$.

The equation of motion for this model is
\be
\frac{d^2\phi}{dx^2}=-n\phi\left(1-\phi^{\frac2n} \right)\left[(n+2)\phi^{\frac2n}-n \right],
\ee
and the kink solution in the sector $0\leq\phi\leq1$ can be obtained analytically; it is
\be\label{solanalytic}
\phi_n(x)=\left(\frac{1+\tanh(x-x_0)}{2}\right)^{\frac n2}
\ee
where $x_0$ is a constant of integration. We have chosen $x_0=\text{arctanh}[(2-n)/(2+n)]$ to make $\phi_n(0)=\phi^+_{m_n}$ as in Eq.~\eqref{phimaxanalytic}. The solution for the sector $-1\leq\phi\leq0$ can be found by taking $\phi_n(x)\to-\phi_n(x)$; in this sector,
$\phi_n(0) = \phi^-_{m_n}$. The kink solution in Eq.~\eqref{solanalytic} has the following limit when $n\to\infty$:
\be
\phi_\infty(x)=e^{-e^{-2x}}.
\ee
It shows explicitly how fast the solution goes to zero for negative values of $x$. The energy density is
\be
\rho_n(x)=\frac{n^2}{2^{n+2}}\left[1+\tanh(x-x_0) \right]^n [1-\tanh(x-x_0)]^2,
\ee
and for $n\to\infty$ one gets $\rho_\infty(x) = 4e^{-4x}e^{-2e^{-2x}}$. 
The energy can be calculated analytically, giving
\be
E_n=\frac12 \frac{n}{n+1}.
\ee
In the limit $n\to\infty$, it gives $E_\infty=1/2$. 

Furthermore, we calculate the portion of energy $E_{0,\phi_{m_n}}$ in the interval $0\leq\phi\leq \phi_{m_n}$, and $E_{\phi_{m_n},1}$ in the interval $\phi_{m_n}\leq\phi\leq 1$. They are
\bes
\be
E_{0,\phi_{m_n}}=\frac12\frac{3n+2}{n+1}\left(\frac{n}{n+2} \right)^{n+1},
\ee
and
\be
E_{\phi_{m_n},1}=\frac12\frac{n}{n+1} - \frac12\frac{3n+2}{n+1}\left(\frac{n}{n+2}\right)^{n+1}.
\ee
\ees
As expected, $E_{0,\phi_{m_n}} + E_{\phi_{m_n},1} = E_n$. Also, the ratio $E_{0,\phi_{m_n}}/E_{\phi_{m_n},1}\to 3/(e^2-3)$ when $n\to \infty$.

\begin{figure}[t]
\includegraphics[width=4.2cm]{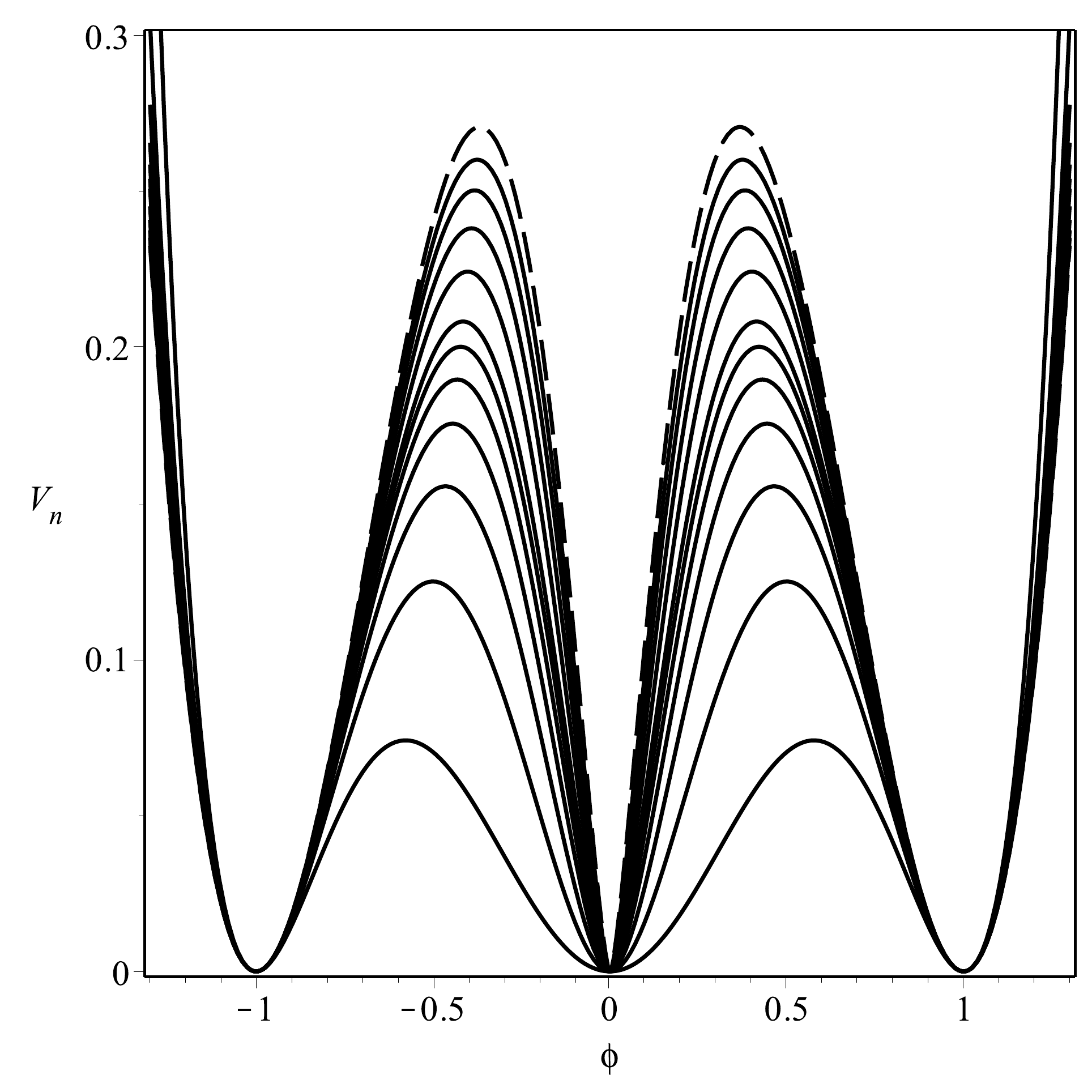}
\includegraphics[width=4.2cm]{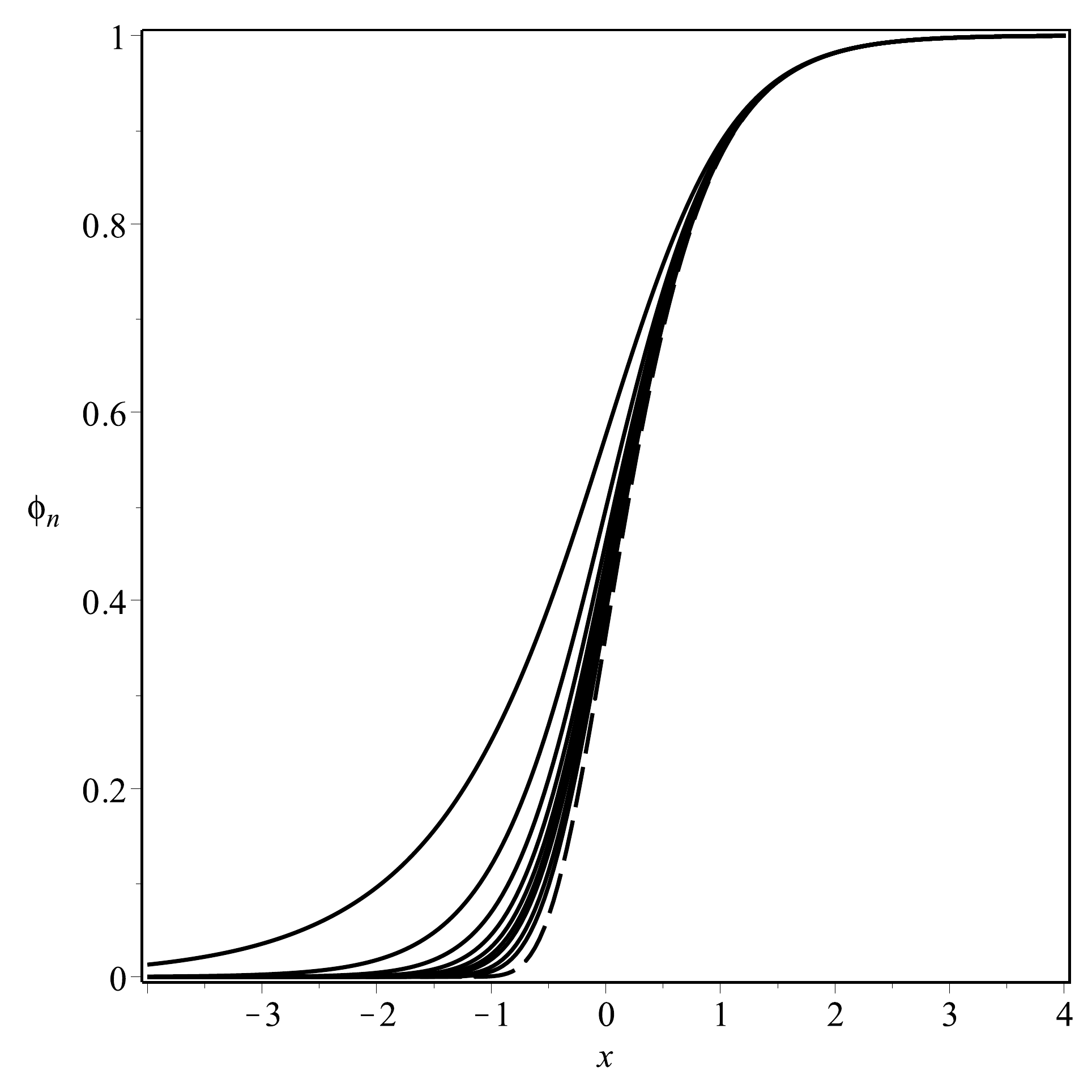}
\includegraphics[width=4.2cm]{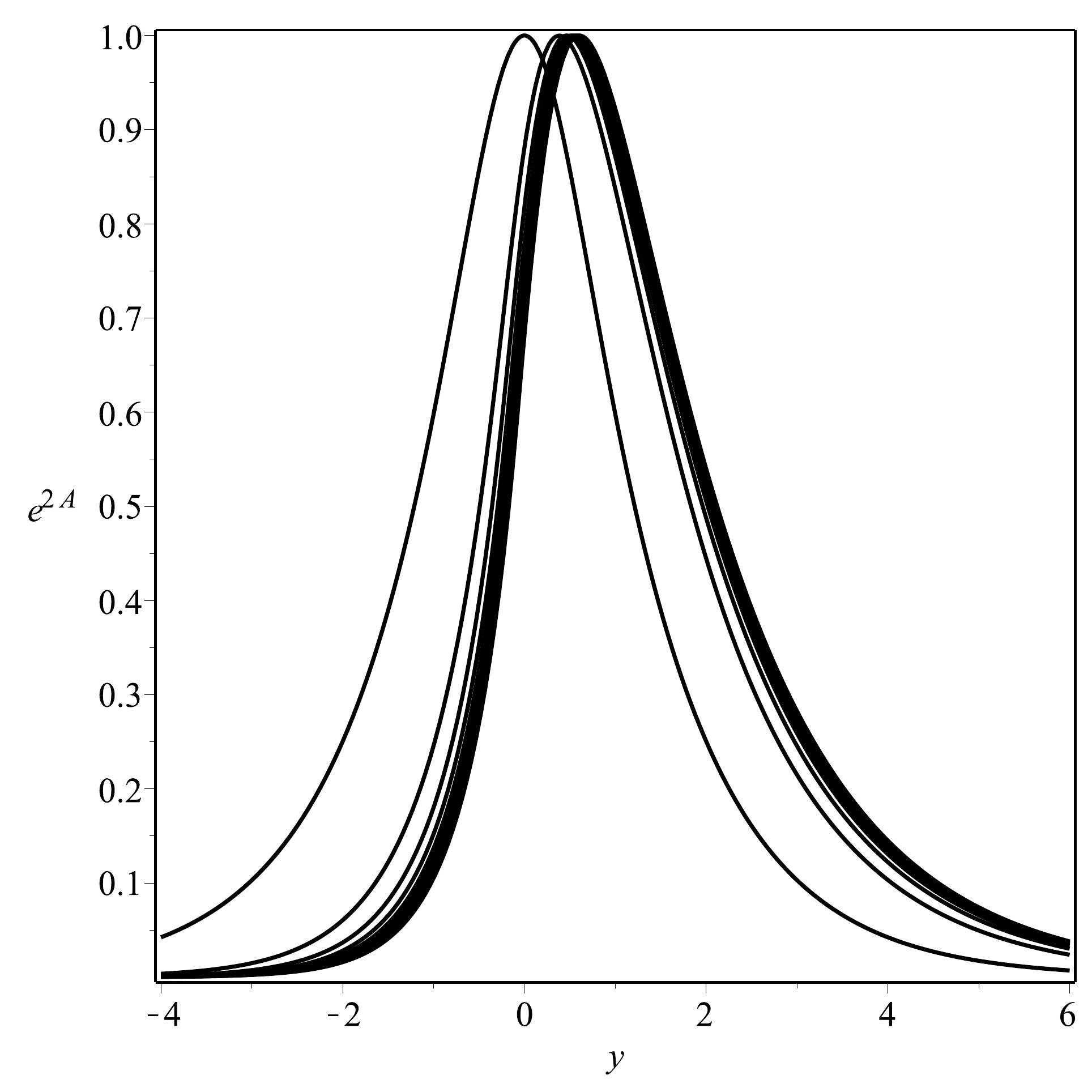}
\includegraphics[width=4.2cm]{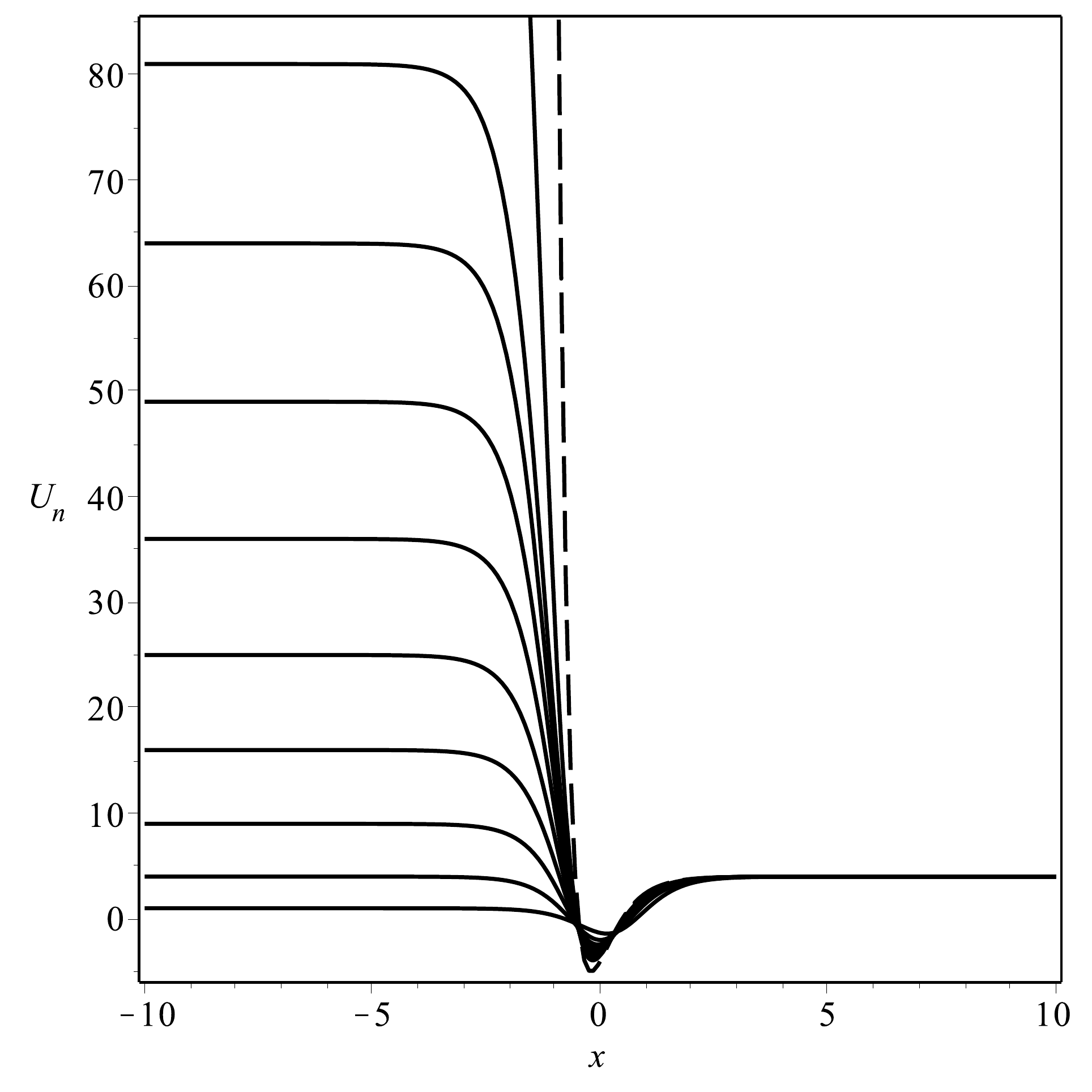}
\caption{The potential \eqref{potphi6like} (top left), kink solution (top right), energy density (bottom left) and stability potential (bottom right), depicted for $n=1$ and increasing to larger and larger values. The dashed lines represent the limit $n\to\infty$.}
\label{fig5}
\end{figure}

The stability potential can be calculated analytically too; it has the form
\ben
U_n(x)&=&\left(2 + \frac{3n}{2} + \frac{n^2}{4} \right)\tanh^2(x-x_0) \nonumber\\
&& + \left(2-\frac{n^2}{2} \right)\tanh(x-x_0) -\frac{3n}{2} + \frac{n^2}{4}.\;\;\;
\een
In the limit $n\to\infty$ one gets $U_\infty(x) = 4-12e^{-2x}+4e^{-4x}$. In Fig.~\ref{fig5} we plot the potential, solution, energy density and the stability potential for several values of $n$.

\section{Asymmetric hybrid brane} 
\label{sec:brane}

We now embed the scalar field action into an warped geometry with a single extra dimension of infinite extent which is described by the line element
\be\label{met}
ds^2_5=e^{2A}\,\eta_{\mu\nu}dx^\mu dx^{\nu}-dy^2,
\ee
with $A=A(y)$ being the warp function, $\eta_{\mu\nu}$ describing the four-dimensional ($\mu,\nu=0,1,2,3$) Minkowski spacetime and $y$ standing for the extra dimension. In this case, the Einstein-Hilbert action has the form
\be
I=\int d^4x dy\,\sqrt{|g|} \left(-\frac14 R+{\cal L}(\phi,\partial_a \phi)\right),
\ee
where $R$ is the Ricci scalar and ${\cal L}(\phi,\partial_a\phi)$ describes the scalar field. Here $a,b=0,...,4$ and we are also using $4\pi G_{5}=1$, for simplicity.

\subsection{Brane model}

We illustrate this case with the potential of Eq.~(\ref{pot2}), but now with the flat brane metric (\ref{met}). We suppose that the scalar field only depends on the extra dimension, $\phi=\phi(y)$. In this case, the equation of motion has the form 
\bes\ben
\frac{d^2\phi}{dy^2}+4\frac{d\phi}{dy} \frac{dA}{dy} =\frac{dV}{d\phi}.
\een
Also, the several Einstein's equations reduce to 
\ben
\frac{d^2A}{dy^2}&=&-\frac23 \left(\frac{d\phi}{dy}\right)^2, \\
\left(\frac{dA}{dy}\right)^2&=& \frac{1}{6}\left(\frac{d\phi}{dy}\right)^2-\frac13 V(\phi).
\een
\ees 
We now take 
\be\label{fob}
\frac{dA}{dy}=-\frac23 W(\phi);\;\;\;\;\;\; \frac{d\phi}{dy}= \frac{dW}{d\phi}.
\ee
These first-order equations solve the equations of motion if the potential is written as
\be\label{pot3}
V(\phi)=\frac12 W_\phi^2 - \frac43 \,W^2.
\ee
The presence of first-order equations ensures linear stability of the gravity sector, so the asymmetric thick brane scenario is robust against fluctuations in the metric.

We can write the energy density in the form
\be\label{rho}
\rho =e^{2A}\left(W^2_\phi-\frac43\, W^2\right).
\ee
Also, we use $W$ as in (\ref{superpot2}) to get
\ben\label{potphi6likebrane}
V(\phi) &=& \frac12 (1-\phi)^2\left(1+\phi^p\right)^2 \nonumber\\
&&-\frac43\left(\phi -\frac{\phi^2}{2} + \frac{\phi^{p+1}}{p+1} - \frac{\phi^{2+p}}{2+p}\right)^2.
\een

\begin{figure}[t]
\includegraphics[width=4.2cm]{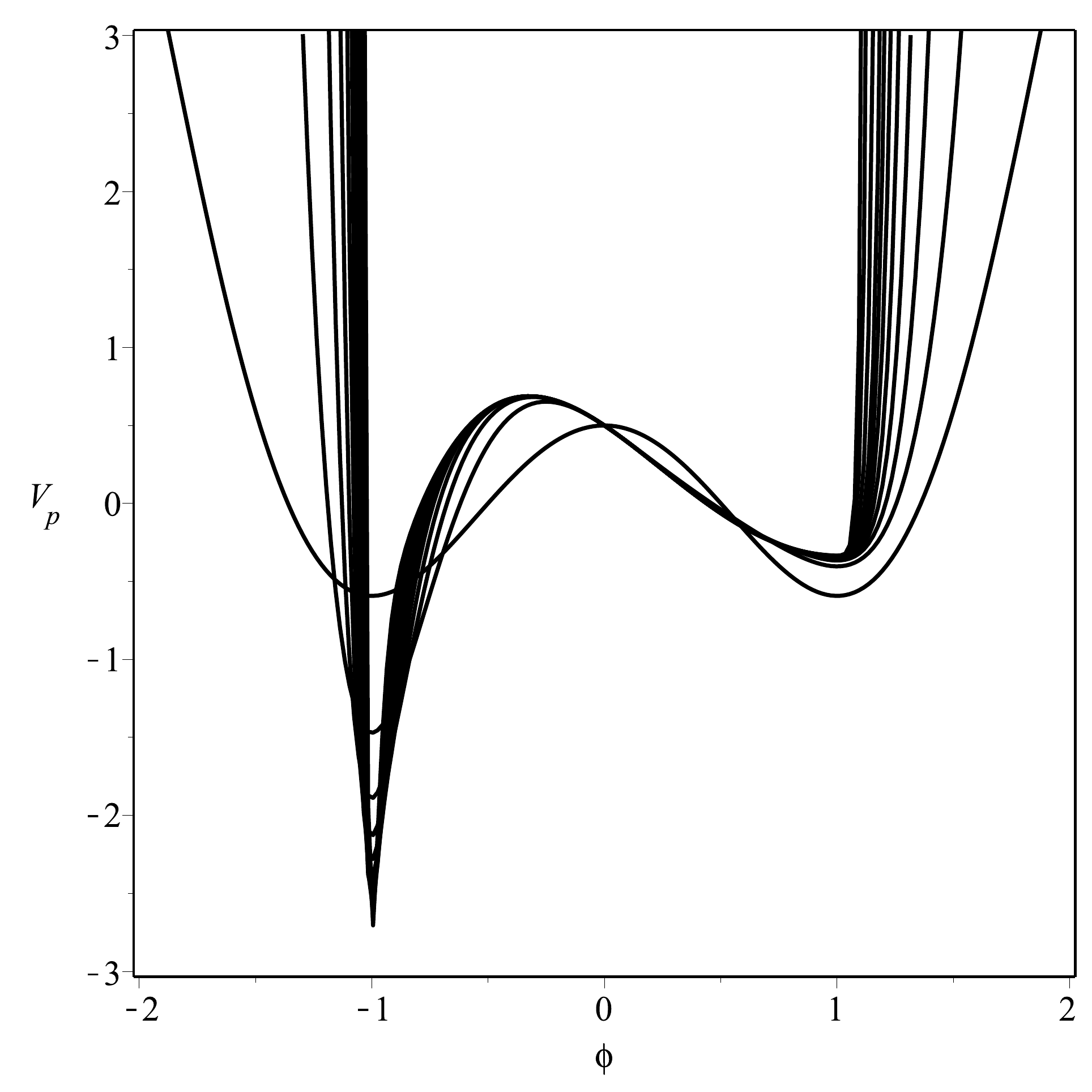}
\includegraphics[width=4.2cm]{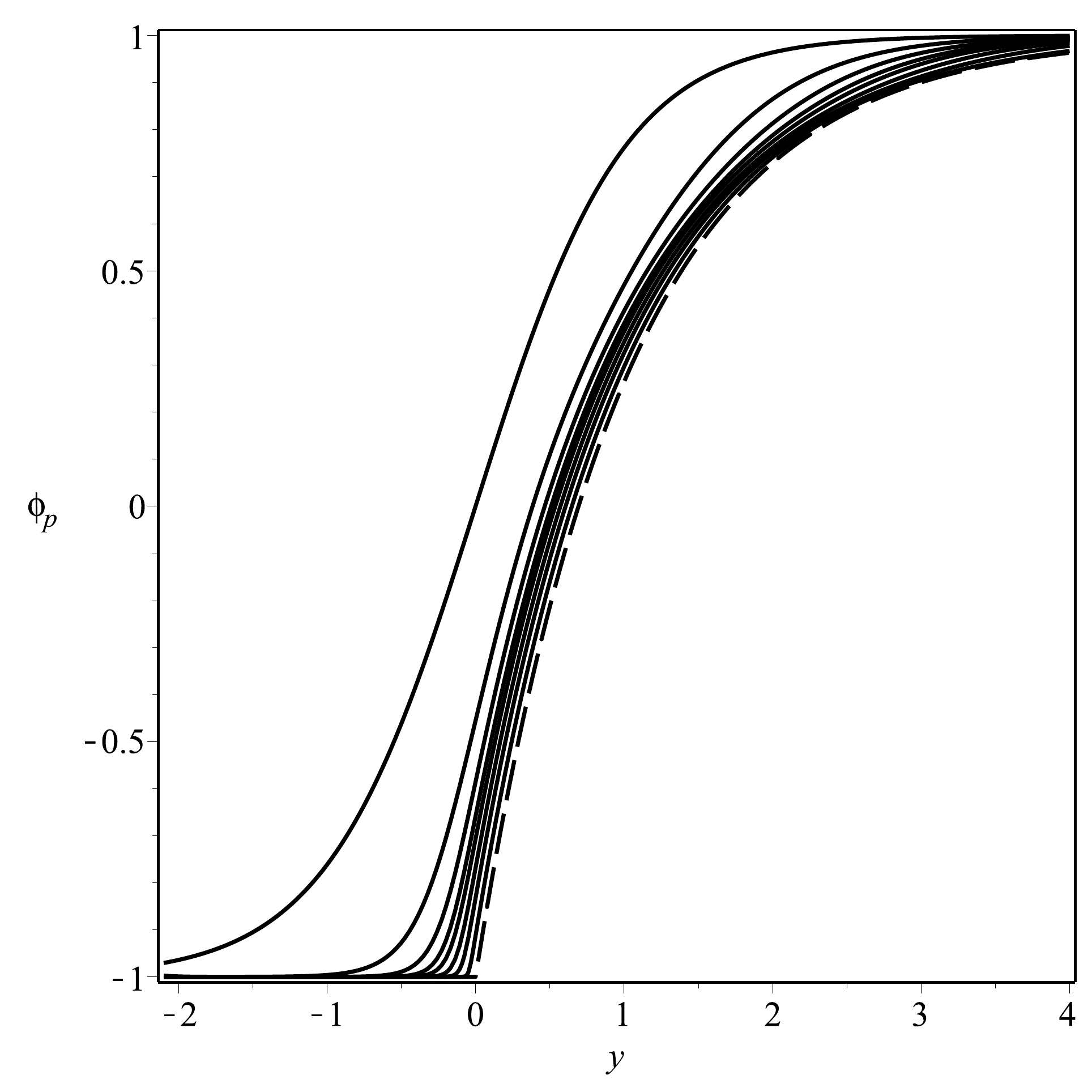}
\includegraphics[width=4.2cm]{fig4c.pdf}
\includegraphics[width=4.2cm]{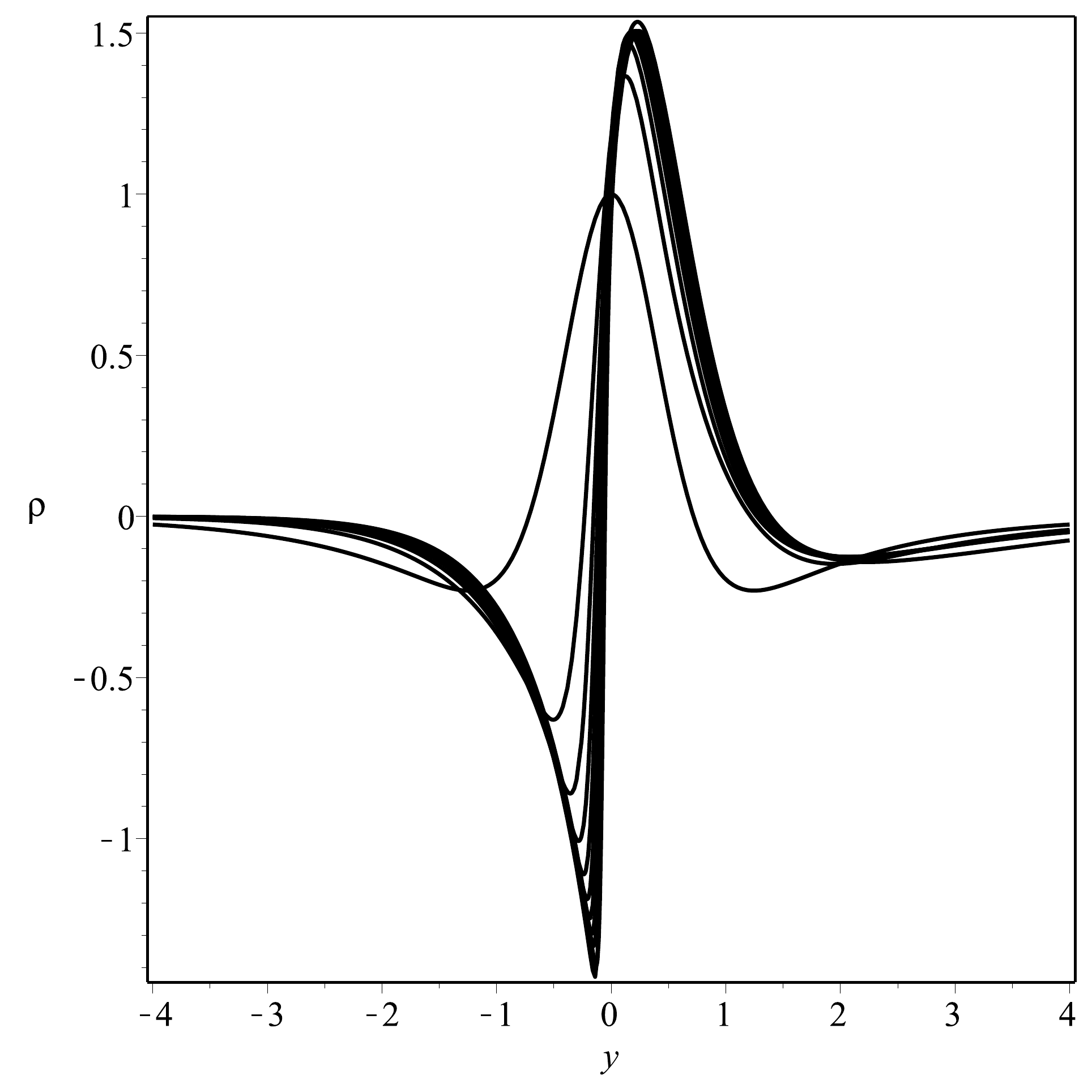}
\caption{The potential (\ref{pot3}) (top left), kink solution (top right), warp factor (bottom left) and energy density (bottom right), depicted for several values of $p$.}
\label{fig6}
\end{figure}

The scalar field solutions are the same of the previous model, studied in flat spacetime. We solve the first-order equations numerically, and we depict in Fig.~\ref{fig6} the potential, kinklike solution, warp factor and the energy density of the asymmetric hybrid brane, for several values of $p$. As we have checked, the potential, solution, warp factor and energy density are all continuous and obey the corresponding equations appropriately.

For $p$ very large, in the limit where the kink becomes a half-compact structure, we use Eq.~(\ref{fob}) to show that the warp function has the form
\be
A(y)=
\begin{cases}
y -\frac23, \,\,\, & y< 0,\\
-\frac13 y -\frac23 e^{-2y}, \,\,\,& y\geq 0,
\end{cases}
\ee 
which is continuous at $y=0$. In the same limit, the warp function can be used to write the energy density analytically, as
\be
\rho(y)=e^{2A(y)}
\begin{cases}
1 , \,\,\, & y<0\\
-\frac13\left(1 - 20e^{-2y} + 16 e^{-4y}\right) ,\,\,\,&y\geq0,
\end{cases}
\ee
which is also continuous at $y=0$. We see that the brane is asymmetric and hybrid, since it behaves as a thin or thick brane, depending on $y$ being at its left or right side.

It is interesting to note from the first-order equations (\ref{fob}) that the energy density (\ref{rho}) can be written as a total derivative, in the form
\be 
\rho=\frac{d}{dy}\left( e^{2A}\,W\right).
\ee
This means that the energy density integrates to zero, leading to an asymmetric thick braneworld scenario with zero energy, despite the asymmetric and somehow exotic form of the energy density which is shown in the bottom right panel of Fig.~\ref{fig6}.

\section{Summary} 
\label{sec:end}

In this work we studied the presence of asymmetric kinklike structures in flat spacetime and its embedding in a warped geometry with a single extra dimension of infinite extent, to generate asymmetric hybrid brane configurations. The current investigation extends the previous investigation \cite{hb}, which describes the presence of symmetric hybrid brane, due to the parity or $Z_2$ symmetry engendered by the source scalar field model. Here we have considered scenarios where the $Z_2$ symmetry is not effective anymore.

As a natural continuation of the current study, one can investigate how to get asymmetric compactlike solutions in a model with kinematics modified to accommodate higher order power in the derivative of the scalar field, and how it behaves embedded in the $AdS_5$ geometry with a single extra dimension of infinite extent. 

There are several other issues of current interest, some of them related to the presence of fermions and gauge fields, to see how they can be trapped inside the hybrid brane, and how the asymmetric features can contribute to the trapping. Another study concerns investigating cosmic evolution using the proposed asymmetric braneworld scenario, inspired from \cite{a2,a6,a7,a8} or taking the alternative route studied in \cite{brito}. Moreover, since we have shown that the source scalar field model may be described by first-order differential equations, one can also wander if the hybrid brane configuration can be extended to include supersymmetry. In particular, one could also study the case of bent brane \cite{bent1,bent2} and ask whether it is possible to build an Hamilton-Jacobi description \cite{st,t} for the novel hybrid brane scenario proposed in this work.

\acknowledgments{The authors acknowledge CNPq for partial financial support: DB thanks support from projects 455931/2014-3 and 06614/2014-6, MAM thanks support from project 140735/2015-1, and RM thanks support from projects 508177/2010-3 and 455619/2014-0.}


\end{document}